\documentclass[preprint2]{aastex}

\usepackage{graphicx}
\usepackage{psfig}

\newcommand{\msun}{\,\hbox{M$_{\odot}$}}
\newcommand{\kms}{\,\hbox{\hbox{km}\,\hbox{s}$^{-1}$}}
\newcommand{\vrot}{\,\hbox{$V_{\rm rot}$}}
\newcommand{\mbh}{\,\hbox{M$_{\rm BH}$}}
\newcommand{\msigma}{\,\hbox{M$_{\rm BH}-\sigma$}}

\newcommand{\degree}{\ensuremath{^\circ}}

\shorttitle{Mass ratio conditions for ULIRG activity in interacting pairs.}
\shortauthors{Dasyra et al.}

\begin{document}

\title{Dynamical properties of Ultraluminous Infrared Galaxies I:
Mass ratio conditions for ULIRG activity in interacting pairs}

\author{K. M. Dasyra, L. J. Tacconi, R. I. Davies, R. Genzel, D. Lutz,
}
\affil{Max-Planck-Institut f\"ur extraterrestrische Physik,
Postfach 1312, 85741, Garching, Germany}
\email{dasyra@mpe.mpg.de,linda@mpe.mpg.de,davies@mpe.mpg.de,genzel@mpe.mpg.de,lutz@mpe.mpg.de}

\author{T. Naab, A. Burkert,}
\affil{University Observatory Munich, Scheinerstrasse 1, 81679, Munich, 
Germany}
\email{naab@usm.uni-muenchen.de,burkert@usm.uni-muenchen.de}

\author{S. Veilleux, }
\affil{Department of Astronomy, University of Maryland, College Park, MD
20742, USA}
\email{veilleux@astro.umd.edu}

\and
\author{D. B. Sanders}
\affil{Institute for Astronomy, University of Hawaii, 2680 Woodlawn Drive,
Honolulu, HI 96822, USA}
\email{sanders@ifa.hawaii.edu}


\begin{abstract}

We present first results from our Very Large Telescope large program to study 
the dynamical evolution of Ultraluminous Infrared Galaxies (ULIRGs), which are 
the products of mergers of gas-rich galaxies. The full data set consists 
of high resolution, long-slit, H- and K-band spectra of 38 ULIRGs and 12 QSOs 
(between 0.042$<z<$0.268). In this paper, we present the sources that have
not fully coalesced, and therefore have two distinct nuclei. This sub-sample 
consists of 21 ULIRGs, the nuclear separation of which varies between 1.6 
and 23.3 kpc. From the CO bandheads that appear in our spectra, we extract 
the stellar velocity dispersion, $\sigma$, and the rotational velocity, \vrot.
The stellar dispersion equals 142\kms\ on average, while \vrot\ is often of 
the same order. We combine our spectroscopic results with high-resolution 
infrared (IR) imaging data to study the conditions for ULIRG activity in 
interacting 
pairs. We find that the majority of ULIRGs are triggered by almost equal-mass
major mergers of 1.5:1 average ratio. Less frequently, 3:1 encounters are
also observed in our sample. However, less violent mergers of mass ratio 
$>$3:1 typically do not force enough gas into the center to generate ULIRG 
luminosities.
\end{abstract}



\section{Introduction}
\label{sec:intro}

In hierarchical cold dark matter models of galaxy formation and evolution,
galaxy merging may lead to the formation of elliptical galaxies,
trigger major starbursts, and account for the formation of supermassive
black holes and quasars (e.g. \citealt{efstathiou}; \citealt{kauffmann}; 
\citealt{haehnelt}). Despite the importance and prevalence of galaxy mergers 
in driving galaxy evolution, the physical details of the merging process are 
not yet well-understood even in the local Universe.

Mergers are responsible for producing some of the most luminous objects of the 
local Universe, the ultraluminous infrared galaxies (ULIRGs). The bolometric 
luminosities of ULIRGs are greater than $10^{12} L_{\odot}$ and emerge mainly 
in the far-infrared (FIR).  ULIRGs are mergers of gas-rich, disk galaxies 
and have large molecular gas concentrations in their central kpc regions 
(e.g. Downes \& Solomon 1998; \citealt{brysco}) with gas-mass densities 
comparable to stellar densities in ellipticals.  

The ULIRG phase occurs in mergers after the first 
peri-passage (e.g. Sanders \& Mirabel 1996, Veilleux, Kim \& Sanders 2002) to 
post-coalescence. The nuclear separation, the presence of tidal tails and 
the high IR luminosities of these sources are all indications that ULIRG 
mergers are in a phase beyond the first approach of the halos (e.g. Veilleux, 
Kim \& Sanders, 2002). These observations are consistent with the results 
from a plethora of numerical models in the literature (e.g. Mihos 
1999; Mihos \& Hernquist 1996; Springel et al. 2005), which indicate that 
starbursts intense enough to drive a ULIRG phase occur only after the 
first encounter and can be present after the nuclear coalescence, before 
complete relaxation sets in.

A quantitative observational technique to investigate galaxy merger evolution 
is to determine the 
kinematic and structural properties of their hosts at different merger 
timescales. With that goal in mind we have conducted a European Southern
Observatory (ESO) large program\footnote{171.B-0442 (PI Tacconi)}, where we 
performed 
high-resolution near-infrared (NIR) spectroscopy of a large sample of ULIRGs 
spanning a wide range of merger phase and infrared luminosity. This work 
expands on the previous spectroscopic studies of Genzel et al. (2001) and 
Tacconi et al. (2002).

In this paper we focus on binary ULIRG sources; these systems are 
between the first and final encounter phases of a merger, thus they still have 
(at least) two well-separated nuclei. We investigate the mass ratios of the 
galaxies that, when merging, produce ULIRG-like luminosities.  The results 
from the remnants, the sources which have coalesced and show a 
single nucleus in the NIR images, will be presented in a forthcoming 
paper, together with the evolution of the host dynamics and the black hole 
mass during the merger.

This paper is arranged as follows.
After summarizing the observations and describing the data reduction 
method in \S~\ref{sec:obs}, we extract structural parameters of our 
sources in \S~\ref{sec:structure}. The stellar
kinematics of the merging hosts, as derived from our long-slit spectra are 
presented in \S~\ref{sec:bhm}. Using the kinematics, we calculate the 
progenitor mass ratio of the merging galaxies in 
\S~\ref{sec:massratio}. To ensure that the observed mass ratio is
not severely affected by the dynamical heating of the system or projection
effects, we perform simulations that predict the time evolution of the mass 
ratio in \S~\ref{sec:model}. An overview of our results is presented 
in \S~\ref{sec:conc}.


\section{Observations and Data Reduction}
\label{sec:obs}

We present near-infrared Very Large Telescope (VLT) spectroscopic
data of local mergers. In the current study, 21 ULIRGs are 
presented, 20 of which are binary systems and 1 of which, 
IRAS 00199-7426, may be a multiple merger (Duc et al. 1997; also see 
Appendix A). To these sources, we add 3 binary ULIRGs that have 
already been presented in \cite{genzel01}.  With the presentation 
of the spectroscopy of 23 binary sources in total, we complete the 
part of our sample that deals with sources in a merger state prior to
the coalescence of the individual nuclei.

The entire sample consists of 38 sources and it is largely drawn 
from the combined 1 Jy catalog (Kim \& Sanders 1998), and the southern-ULIRG 
(SULIRG) sample of the Duc et al. (1997) study. One source, IRAS 02364-4751,
is from \cite{rig99}.  The sample size increases to 54 ULIRGs 
when the sources studied in \cite{genzel01} and \cite{tacconi02} are included.
The 1 Jy catalog comprises a complete flux-limited (at 60 $\mu$m) sample of 
118 ULIRGs compiled from a redshift survey of IRAS Faint Source Catalog 
version 2 objects (Moshir et al. 1990). Veilleux et al.~(2002) have completed 
and analyzed an R- and K-band survey of the entire catalog, such that 
photometric and structural data (absolute magnitudes, surface brightnesses, 
half-light radii) are readily available.  We have observed those sources with 
dec $<25\degree$, and with redshifts where the strong rest frame H-band 
stellar absorption lines lie in parts of the H- and K-band with high 
atmospheric transmission (z$\le$0.11 and z$\ge$0.20). 

\begin{figure*}
\centering
\includegraphics[height=5.2cm,width=13.6cm]{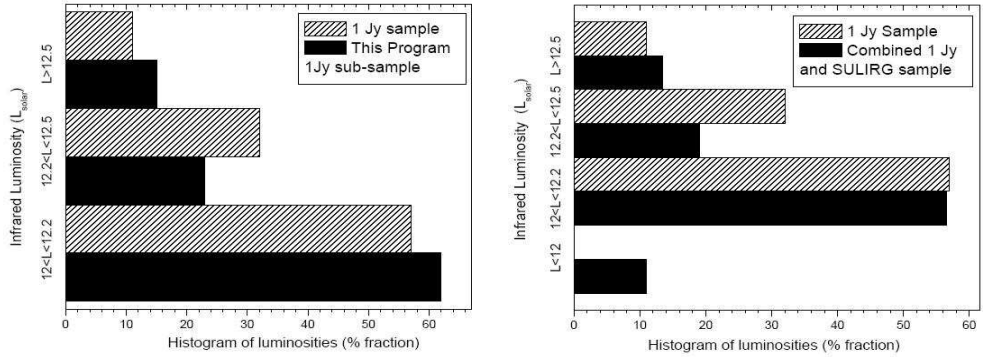}
\caption{ Histogram of luminosities of samples used in this study.
Sources from the  
1 Jy catalog (Kim et al. 2002) are denoted by the hatched bars. In the
left panel, the sources we selected from the 1 Jy catalog are shown as
filled bars and follow well the original sample's luminosity distribution. 
In the right panel, the mean luminosity of the combined samples is reduced
due to the addition of the lower-luminosity \cite{duc97} sample.
\label{fig:sample}}
\end{figure*}

The left-panel histogram of Fig.~\ref{fig:sample} shows that the sources we 
selected from the 1 Jy catalog follow a similar luminosity distribution as
 the entire catalog. Given that the latter is 
solely compiled according to the 60 \micron\ flux, it does not favor 
any particular pre-merger initial conditions.
When adding sources from the \cite{duc97} catalog, which contains less luminous
sources than the 1 Jy sample (see right panel of  Fig.~\ref{fig:sample}), the 
average IR luminosity of our sample is reduced, but remains
luminosity-selected. For the sources of the \cite{duc97} sample, we do not
adopt the $L_{{\rm IR}}$ values of the authors, but we use the \cite{sami96} 
expression and the Faint Source Catalog version 2 mid-infrared (MIR) and FIR 
fluxes to calculate $L_{{\rm IR}}$.
Two of the sources in our large program sample are less luminous than $10^{12} 
L_{\odot}$, however we also treat them as ULIRGs given that the classification 
often depends on the accuracy of the mid- and far-infrared flux measurements.

Our data were taken with the VLT ANTU telescope on Cerro Paranal, Chile. 
We used the ISAAC spectrometer (Moorwood et al. 1998) in mid-resolution mode 
in the H band ($\lambda / \delta \lambda = 5100$), and in the K band 
($\lambda / \delta \lambda = 4400$), with a slit width of 0.6\arcsec.  The 
on-chip integration was 600 s per frame with typical total integration times 
of 1 hr per slit position angle (see Table~\ref{tab:list}). For most of the
binary sources we observed along three slits, with the first slit
going through both nuclei. The other two slits are (usually) 
perpendicular to the first one and go through the brighter and the fainter 
nucleus respectively.

We have selected the central wavelength in a way 
such that most of the CO(3-0), CO(4-1), SiI, CO(5-2), and CO(6-3) H-band 
bandheads (at 1.558, 1.578, 1.589, 1.598, and 1.619 \micron\ respectively), 
as well as the forbidden [FeII] emission line (at 1.645 \micron), appear in our
spectra. For the sources with redshift $z>0.2$ we use (some of) the CO(8-5), 
CO(9-6), and CO(10-7) absorption bandheads (at 1.661, 1.684, and 1.706 
\micron\ respectively), which are then shifted to the K band. The CO and
SiI absorption features trace the stellar, while the Fe emission 
line traces the warm gas kinematics. The observed central wavelength range 
varied from 1.68 to 2.08 \micron, depending on the redshift of each source 
(Table~\ref{tab:list}). The most nearby of the objects presented in
this study is at redshift $z=0.0431$ while the most distant at $z=0.242$.   

For the data reduction we used standard IRAF routines.  We first subtracted the 
frames of positive from the frames of negative chop throw (offset from the 
telescope pointing position) for 
the sky background removal, and flat-fielded the result. Then, we performed 
a bad-pixel and cosmic-ray removal, and corrected for detector deformations. 
For the spatial direction, we combined several spectroscopic frames of a 
point-like source (star) at a different chop throw and nod (random offset, 
smaller than the chop throw). By fitting all the stellar traces, we found the 
low-order polynomial that best corrects for deformations of the spatial 
axis. For the spectral axis we used a ''sky'' frame, which simply 
was a randomly selected, dark-subtracted frame of our exposures. We found the 
best wavelength correction matrix by identifying the sky-lines in that frame,
and, again, by fitting a polynomial to them. After rectifying the images in 
both the spatial and wavelength directions, we spatially shifted the frames
so that their traces overlap and, then, we combined them. The spectral 
extraction from the final frame was followed by an atmospheric correction 
with the aid of a telluric (usually B dwarf or solar type) star. The spectral 
extraction procedure was repeated for several apertures along each slit,
and for two different slit position angles, so that the two dimensional image
of the stellar kinematics could be reconstructed. The final spectra were 
shifted to restframe.

To extract the velocity dispersion $\sigma$ and rotational velocity 
$V_{\rm rot}$ we correlated the source spectra with that of an appropriate
template star. Due to the
starburst nature of a ULIRG, the stellar population that dominates the 
near infrared (NIR) light is either a giant or a supergiant (or a combination 
of the two). For this purpose we selected either HD 25472 or HD 99817
(M0III giant and M1I supergiant respectively). We used the Fourier correlation
quotient (FCQ) technique described in \cite{bender90} with a Wiener filter to 
suppress the high-frequency noise; for this we used a code written by  
one of us. The FCQ technique is based on the
deconvolution of the correlation function peaks of the source and the stellar 
template to the autocorrelation function peaks of the template. It provides
the broadening function along the line-of-sight (LOS) of the observations. 
We fit a high-order Gaussian (linear combination of Gaussian and second 
order polynomial) to the broadening function in order to derive 
the stellar dispersion and the recession velocity, $V_{\rm rec}$. 
For this purpose, we use all of the above-mentioned H-band bandheads that
exist in our spectra, as long as the signal-to-noise allows us to do 
so, and we average the results.
From the difference in the recession velocity along several apertures of the 
slit, we calculate the rotational velocity on the plane defined by the 
line-of-sight and the position angle of the slit.

\begin{figure*}
\includegraphics[height=20cm,width=16cm]{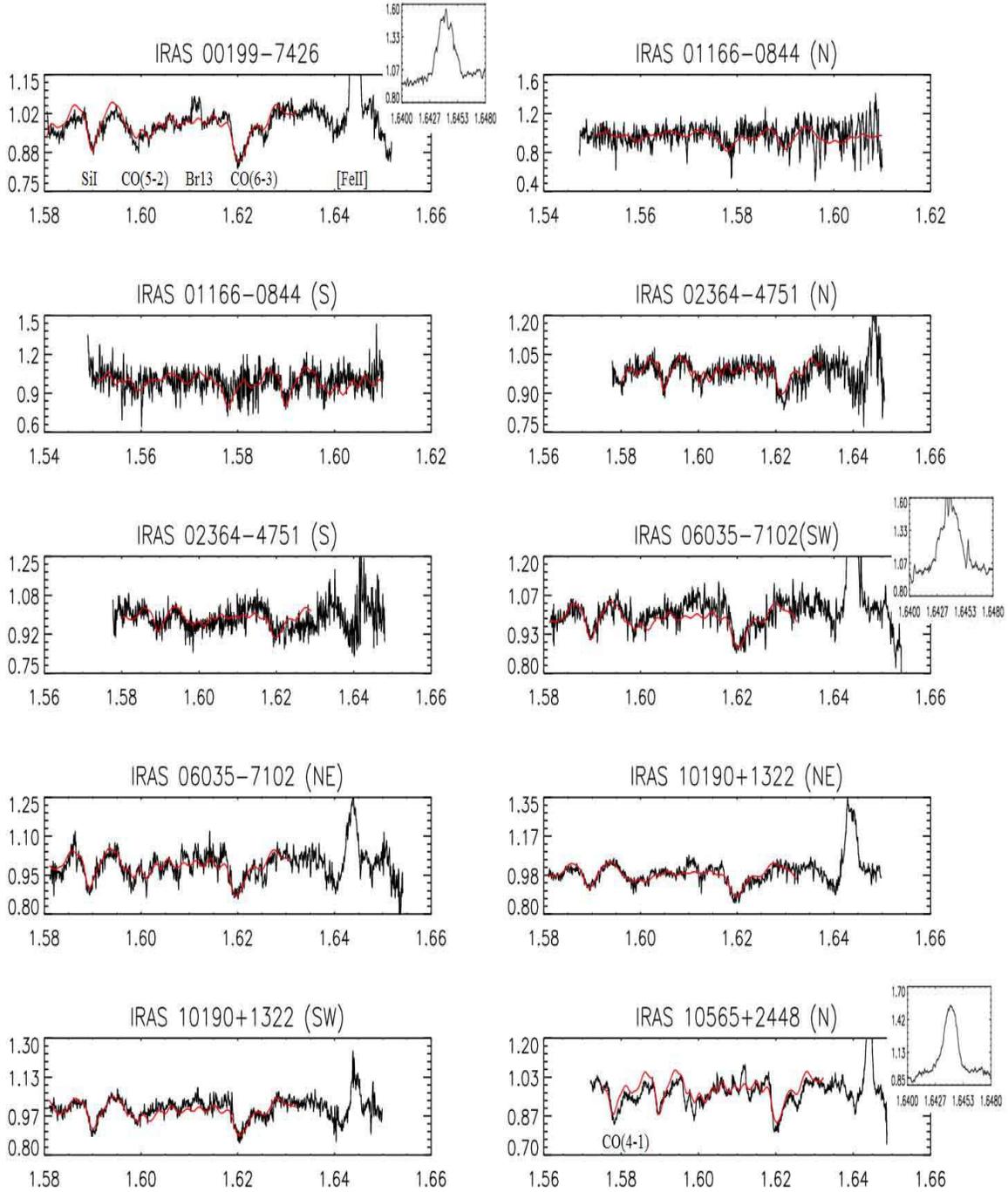}
\begin{center}
\caption{The reduced H-band spectra of the binary ULIRGs. 
The stellar templates, convolved with a Gaussian that represents 
their LOS broadening function,  are overplotted in solid line.
All the spectra are shifted to rest frame. \label{fig:spectra} }
\end{center}
\end{figure*}

We follow the above procedure to extract the spectra for each source 
(or nucleus). The central aperture spectra, combined over the slits and 
shifted to the restframe, are displayed in Fig.~\ref{fig:spectra}. In each
panel, the stellar template is overplotted with a solid line, after being 
convolved with the Gaussian that best fits the respective LOS broadening 
function.


\section{Structural parameters}
\label{sec:structure}

The conversion of our dynamical measurements into masses requires 
complementary data that trace the structure of our sources, namely the
half-light-radius $R_{\rm eff}$ and the inclination to the line of sight, 
$i$. 
  
Given that ULIRGs originate from the merger of gas-rich disk galaxies, we use the 
(dynamically perturbed) progenitor disks to estimate the inclination. The rotational 
velocity of a disk 
is connected to its line of sight dependent value, $V_{\rm LOS}$ as follows  
\begin{equation}
\label{eq:1}
\mbox{$V_{\rm rot}=V_{\rm LOS}/ (cos(\phi_{\alpha}) sin(i))$}.
\end{equation}
The parameter $\phi_{\alpha}$ is the angle between the slit position angle 
and the major axis of the inclined disk (which is an ellipsoid when projected 
in 2 dimensions). 

\begin{figure*}
\includegraphics[width=16cm]{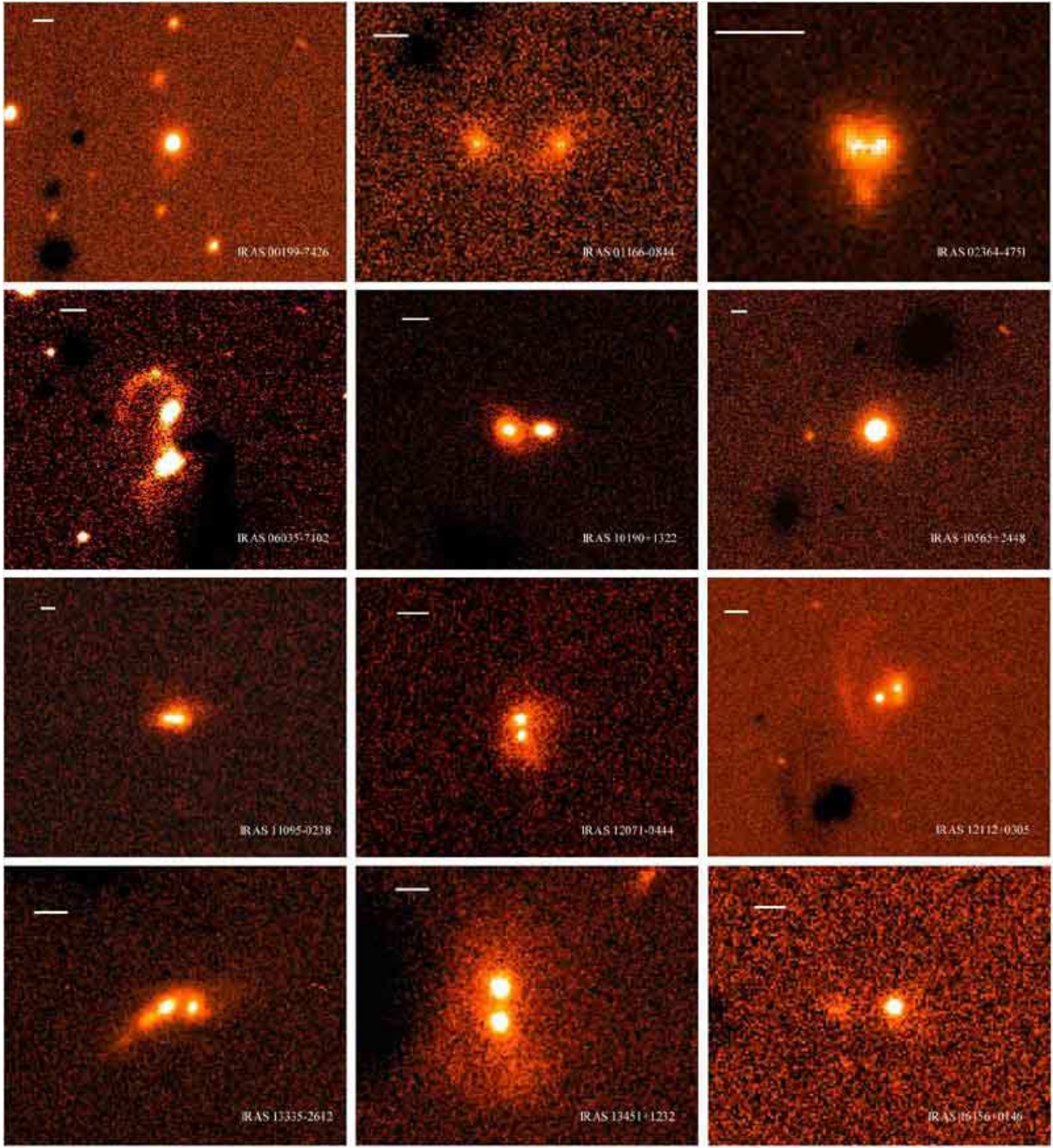}
\begin{center}
\caption{The (raw) H-band acquisition images. The horizontal line in the 
upper left corner of each panel corresponds to 5 kpc at the reshift of the 
source. 
\label{fig:acquisition} }
\end{center}
\end{figure*}

We derive the structural parameters $i$ and $\phi_{\alpha}$ for the stellar
disk of each ULIRG by fitting ellipses to the H-band acquisition images
(see Fig.~\ref{fig:acquisition}).
The fit is performed with the aid of the SExtractor package 
(\citealt{sextractor}), made available by the Institut d'Astrophysique de
Paris. We first detect the center and the radial extent of each source by 
setting a threshold that separates the sky background from any real
detection. We then deblend sources that spatially overlap to obtain the 
apparent ellipticity $\epsilon$, the angle  $\phi_{\alpha}$ (which appear in 
Table~\ref{tab:structure}), and the enclosed counts of each ellipsoid.

The apparent ellipticity is related with the inclination $i$ of the heated
stellar disk as
\begin{equation}
\label{eq:2} 
\mbox{$\epsilon (2-\epsilon)=\epsilon_t (2-\epsilon_t) (sini)^2$}
\end{equation}
(\citealt{binneybook}; Chapter 4.3). The quantity $\epsilon_t$ is the 
(true) ellipticity of the heated disk when seen edge-on. We assume that the 
ratio of the thickness to the truncation radius is 0.3 for the binary ULIRGs, 
which is the 
average value between field spirals and disky ellipticals (\citealt{binney81}).
In this case $\epsilon_t$ equals 0.7. The inclinations calculated with this
method are presented in Table~\ref{tab:structure} and have a mean value
of 43\degree. We note that when using the flat disk approximation 
($\epsilon_t=1$) the mean inclination of this sample is 40\degree. 
Solving and differentiating Eq.~[\ref{eq:2}] for $i$ shows that the smaller 
the inclination, the greater the error on its measured value for a given 
$\epsilon$. The systems that are close to face-on are, therefore, those with 
the most uncertain inclination estimates.

We use the half-light radius as the fiducial aperture in which to calculate 
masses and luminosities for the progenitor nuclei.  However, the half-light
radii for most of the individual nuclei of our binary ULIRGs are not 
readily available in the literature; several binary systems have been treated 
as a single object (e.g. \citealt{veilleux02}; Scoville et al. 2000), often 
due to low angular resolution. When available, the  effective radii are not 
usually measured from NIR data but, from optical bands where the light 
extinction is significant. Due to the extremely dusty environment of ULIRGS 
and to inclination effects, average extinction corrections are not always 
reliable for individual sources. For these reasons, we measure new half-light 
radii from our H-band acquisition images by fitting ellipsoids to the 
individual nuclei  and finding 
the radius at which the ellipsoid contains half of the total counts.
We tabulate the measured H-band $R_{\rm eff}$ in Table~\ref{tab:structure},
after converting angular distances into linear sizes. All distances
in this paper are for a H$_0$=70 km \hbox{{s}$^{-1}$} \hbox{{Mpc}$^{-1}$},
$\Omega_{m}$=0.3, $\Omega_{\rm total}$=1 cosmology.

Our results are consistent with those of NIR imaging available in the
literature, despite the fact that the acquisition images have short exposure
times ($\sim$10 s) and could be tracing only the most luminous parts of the 
sources,
leading to underestimates of the true half-light radius. To check this possible
bias, we compare the effective radii for the sources we have in common with
\cite{sco00}. We find that the effective radii for IRAS 12112+0305 (sw), 
IRAS 13451+1232 (w),  and IRAS 22491-1808 (e) are 0.81, 4.14, and 1.99 kpc 
while the half-light radii for flux within 3 kpc given by \cite{sco00} 
(at 1.6 \micron) were 0.79, 1.07, and 1.66 kpc respectively. The results for 
two of the cases are very similar and the disagreement in the case of IRAS 
13451+1232(w) is due to aperture effects. IRAS 13451+1232 is one of the most
extended sources of the \cite{sco00} sample with a radial extent $>7$ kpc.
Furthermore, Veilleux et al. (2006, in preparation) have recently acquired HST 
NICMOS H-band imaging for several ULIRGs of the 1 Jy catalog and have 
performed a two-dimensional decomposition of the AGN point spread function
(PSF) and the host. The effective radii measured from our acquisition images
are in good agreement with those of the PSF-subtracted hosts of Veilleux et
al. (2006, in preparation).

The structure of a merger, and in particular the nuclear separation, can be
used to trace the timescales of each merging system (e.g. Barnes 2001). The 
majority of the pre-merged ULIRGs have intrinsic nuclear separation smaller 
than 10 kpc (see Table~\ref{tab:structure}), a fact that classifies them as 
pre-merger close binaries according to the \cite{surace98} scheme. Only 
five of our galaxies, IRAS 01166-0844, IRAS 06035-7102, IRAS 10565+2448, 
IRAS 19254-7245, and IRAS 21208-0519 
are considered wide binaries in the same classification scheme. The mean 
projected nuclear separation of our sample is 7.3 kpc (and the median 
5.4 kpc). 

\section{ULIRG stellar velocities and black hole masses}
\label{sec:bhm}
The stellar dispersions extracted (according to the prescriptions of 
\S~\ref{sec:obs}) by the Fourier quotient technique from the central-aperture
spectrum of each source are listed in Table~\ref{tab:velocities}. 
In the fainter sources, $\sigma$ may be somewhat overestimated (at most 
by 20\%) due to low signal-to-noise ratio, which can mimic broader 
dispersions.
The stellar velocity dispersion may vary when measured from different 
bandheads (typically by 15\%). This is both due to a possible template 
mismatch and to the sky-line contamination of our spectra. The velocity 
error bars are equal to the standard deviation of the measurements 
performed at the individual bandheads.

The mean observed dispersion of our binary ULIRG sample, combined with 
the sources in \cite{genzel01}, is 142 km s$^{-1}$ (with a standard deviation 
of 21 \kms).
Sources of intrinsic nuclear separation close to or less than 1 kpc (Arp 220 
and NGC 6240, see Genzel et al. 2001, Tecza et al. 2000 ) were removed 
from the above statistics. By the time the nuclei of two merging galaxies 
are separated by $\lesssim$ 1 kpc, the stellar velocities have 
almost reached their final relaxation values (Genzel et al. 2001, Mihos 2000, 
Bendo \& Barnes 2000). As a consequence, these systems have dispersions 
very close to their (common) equilibrium value and resemble more the 
coalesced ULIRGs, despite the fact that their nuclei can still be resolved.

We measure the rotational velocity along each slit and we correct it for 
the angular deviation $\phi_{\alpha}$ from the major axis of rotation
as discussed in \S~\ref{sec:obs}.
After averaging the results over the slits, we obtain the observed 
rotational velocity, $V_{\rm rot}(obs)$, which we display in 
Table~\ref{tab:velocities} together with its error bar (calculated
similarly to that of $\sigma$). In the same Table we also present the 
final, inclination corrected rotational velocity $V_{\rm rot}$. 

The ratio of the observed stellar rotational velocity to the dispersion, 
$V_{\rm rot}{\rm (obs)}/\sigma$,
is given in Table~\ref{tab:velocities} for each source. The mean 
$V_{\rm rot}{\rm (obs)}/\sigma$ ratio for the sample presented in 
this study is 0.42, while when using the inclination corrected velocity,
the ratio $V_{\rm rot}/\sigma$ increases to 0.77.  Both values are low
compared to those of spiral galaxies. We now investigate whether this 
result is due to the violent relaxation 
process or due to systematics, such as beam-smearing effects. To check for 
beam smearing we calculate the $V_{\rm rot}/\sigma$ ratio for the sources 
for which we have been able to derive rotation curves (due to their large 
radial extent). These are the sources with z$<$0.07 as well as IRAS 20046-0623.
We find that the $V_{\rm rot}/\sigma$ ratio for these sources is similar to 
that of our entire sample: 0.58 and 1.16 when using the inclined-disk and 
inclination-corrected velocities, respectively. We conclude that the low 
rotational velocitites observed in the binary ULIRGs is due to the 
actual dynamical 
heating of the merging systems. Similar conclusions are drawn from the
work of \cite{mihos00}, who presents simulations of 
the velocity moments during the merger process. The 
$V_{\rm rot}{\rm (obs)}/\sigma$ ratio implied from \cite{mihos00} for our
median nuclear separation (5 kpc) and for the radius containing 50\% of 
the stellar mass (or the $R_{\rm eff}$ for constant M/L within the galaxy) 
is also $\sim$0.4.

Using the stellar dispersions listed in Table~\ref{tab:velocities}, we  
estimate a BH mass, $M_{BH}$, with the aid of the $M_{BH}-\sigma$ relation 
(e.g. Gebhardt et al. 2001; Ferrarese \& Merritt 2001).
The published estimates for the slope of the \msigma\ relation span a  
significant range (see \citealt{tremaine02}; \citealt{gebhardt01}; 
\citealt{mefe01}). We use the \cite{tremaine02} expression  
\mbox{$M_{BH}=1.35 \times 10^{8} (\sigma/200)^{4.02}$\msun} which lies
between those of \cite{gebhardt01} and \cite{mefe01}.
We present the BH mass calculated for each source in
Table~\ref{tab:velocities}. The mean black hole mass of the binary 
ULIRG sample is an order of 
magnitude greater than that of the Milky Way and equals $3.9 \times 10^7$
\msun\ (for each nucleus). Converting the stellar dispersions into black hole 
masses carries the uncertainty of applying the \msigma\ relation to 
systems that are not in dynamical equilibrium. The errors introduced by
this conversion and the conditions under which the \msigma\ relation may
provide an accurate estimate of \mbh\ during a merger will be presented in a 
forthcoming paper (Dasyra et al. 2006, in preparation). 

In Table~\ref{tab:velocities} we present the (minimum) black hole mass that
each source would have, if it were accreting at the Eddington rate
($L_{{\rm Eddington}} / L_{\odot}=3.8 \times 10^{4} M_{BH}({\rm Eddington}) 
/ M_{\odot}$). We assign to the Eddington luminosity $L_{{\rm Eddington}}$ 
half of that emitted in the IR (Genzel et al. 1998; Sanders \& Mirabel 1996). 
This is a statistically plausible assumption based on the fact that some 
ULIRGs are largely AGN- while others are starburst- powered (see 
\citealt{genzel98}; \citealt{duc97}; Lutz et al. 1999). However, for
individual sources, the numbers given in Table~\ref{tab:velocities} may be 
higher up to a factor 2 or much lower. We assign the luminosity to each 
nucleus according to the K-band luminosity ratios (Kim et al. 2002; 
\citealt{duc97}), under the assumption that both progenitors have
a BH. To distribute the luminosity between the two nuclei of IRAS
12071-0444 and IRAS 21329-2346 we used the H-band count ratios (1.23 and 
2.17 respectively; also see the Appendix) since no photometric information on 
individual nuclei 
was available in the literature. For the sources of apparent nuclear 
separation $<$ 0.7$''$ we distributed 50\% of the luminosity to each nucleus, 
since we used pixel masking (that affects the number counts) to deblend
the progenitors. The ratio of the Eddington to the dynamical BH mass, the
Eddington efficiency $\eta_{{\rm Edd}}$, is given in the last column of 
Table~\ref{tab:velocities}. On average, it is 0.34 for the individual 
nuclei, which implies that at the pre-merger phase the accretion onto the BH 
is lower than the Eddington limit.


\section{Progenitor mass ratios}
\label{sec:massratio}

For the binary ULIRGs presented in this study, the 
stellar kinematics allow us to find the progenitor mass ratios, $r_m$,
using the virial theorem. We assume a King model to relate the  
observed (LOS) dispersion to the total bulge dispersion.   
The disk and gas mass are accounted for by adding the contribution of 
the (inclination corrected) rotational velocity (for the cases where 
the measurement of \vrot\ is possible). The dynamical mass enclosed 
within an effective radius is then proportional to
\begin{equation}
\label{eq:3}
\mbox{$M\propto R_{\rm eff}(3 \sigma ^2+V_{\rm rot}^2)$},
\end{equation}
Further factors that take into 
account the galactic structure are not important here since we are only 
interested in the mass ratio of the merging systems. We use the values of 
\vrot, and $\sigma$ of Table~\ref{tab:velocities}, and the half-light radius 
of Table~\ref{tab:structure}.

We present the bulge mass ratio $r_m({\rm bulge})$ (calculated only using the
dispersion velocity) and the total baryonic mass ratio $r_m$ (calculated using 
both $\sigma$ and \vrot) in the first two columns of Table~\ref{tab:ratio}.
For both ratios, the mass enclosed within the effective radius of each 
progenitor was used. The convention we use in this Table is that the total 
mass ratio $r_m$ is greater than unity. As a consequence, a $r_m({\rm bulge})$ 
value $<1$ means that the more massive galaxy of the pair has the less 
massive bulge.

For the sources that have a \vrot\ 
measurement, the mean mass ratio equals to 1.40 when only the 
bulge is considered, and 1.35 when the stellar disk is added. As a
consequence, the difference between using the bulge and the total 
baryonic ratio is so small that it allows us to safely use the former
for the cases where we were not able to extract \vrot.

The mean progenitor mass ratio derived from Table~\ref{tab:ratio} is 
1.54, and shows that the majority of the sources we studied are major mergers
of 1:1 to 2:1 progenitor mass ratios. As major mergers we denote systems
of progenitor mass ratio as high as 3:1; mergers of 4:1 or greater mass ratio
are considered minor.
Progenitors of apparent nuclear separation less than 0.7$''$ overlap on the 
detector (even though their nuclei are resolved), because they are spatially
extended (with an average sample FWHM of 5 pixels). Their measured kinematics
depend on the kinematics of their counterparts and, thus, we have decided not
to include them in our statistics (IRAS 02364-4751, IRAS 11095-0238). Our 
result is in agreement with several merger models in the 
literature (e.g. ~\citealt{mihos94};~\citealt{naab03}) that attribute
the ultra-luminous phase to major mergers. 

Another indication of the progenitor mass ratio can be drawn from the 
remnant $V_{\rm rot}{\rm (obs)}/\sigma$ ratio.
\cite{naab03} performed gas-free, N-body simulations of binary mergers 
of several mass ratios and orientations, and found that the major mergers 
were those that led to slowly rotating remnants. They suggested that the 
$V_{\rm rot}{\rm (obs)}/\sigma$ ratio is $\sim 0.2$ for 1:1 and $\sim 0.4$ 
for 2:1 merger remnants, while it reaches higher values (0.8) for minor
merger remnants. The  $V_{\rm rot}{\rm (obs)}/\sigma$ ratio for the merged
ULIRGs of this study, which will be presented in a future paper (Dasyra et al. 
2006, in preparation), is in good agreement with the results of \cite{naab03} 
and the more recent simulations of \cite{burkert05}.

Further observational evidence for the mass ratio of sources with luminosity
cutoff $>10^{12} L_{\odot}$ comes from the work of Ishida (2004), who
calculated the optical (B-band) luminosity ratio of Luminous Infrared Galaxies 
(LIRGs), sources of $10^{11} L_{\odot}< L_{\rm IR} <10^{12} L_{\odot}$. Ishida 
(2004) found a trend of decreasing luminosity ratio with increasing 
luminosity cutoff. Interacting sources of  
$L_{\rm IR} <10^{11.5} L_{\odot}$ were characterized by a wide spread
in the optical luminosity ratio. However, the majority ($>80$\%) of sources of 
$L_{\rm IR} >10^{11.5} L_{\odot}$ were strongly interacting (wide binaries
with disturbed morphologies, tidal tails or internuclei bridges) or merging 
pairs with luminosity ratios $<$ 4:1. Since at least a sub-sample of the 
high-luminosity LIRGs will likely evolve into ULIRGs, the ULIRG luminosity 
ratios should be expected to have similar (or smaller) luminosity ratios.

On the other hand, we do not exclude the possibility of a minor merger 
evolving into a ULIRG. IRAS 20046-0623 does show a second nucleus in both 
the H- and R-band images, which is however too faint to be deblended from 
the bright source or to be spectroscopically reduced. IRAS 10565+2448, has
an H-band luminosity ratio (calculated from the acquisition image) which is 
consistent with a 5:1 merger. Due to extinction effects, imaging results are 
not as reliable as spectroscopic ones in the tracing of the system mass, so 
this 5:1 ratio is only an indication that, even rarely, minor mergers may 
appear in our sample.

To address whether the luminosity can actually trace the mass of these dusty
systems, we compare the luminosity ratio to the mass ratio of our ULIRGs. In 
the literature, the luminosity ratio is calculated within a specific aperture, 
equal for both nuclei. To be consistent in our comparison we also need to 
calculate the baryonic mass ratio inside  a given aperture, which we name 
$r_m({\rm aperture})$. We present the latter ratio and the size of the 
selected aperture in Table~\ref{tab:ratio}. The use of a common aperture 
for both progenitors instead of their effective radii can make the 
intrinsically fainter nucleus appear brighter than its counterpart (IRAS 
06035-7102, IRAS 10190+1322, IRAS 11095-0238, IRAS 21130-4446, IRAS 
12112+0305). The area that is used for the calculation of the mass ratio may 
also significantly change the results.

\begin{figure*}
\centering
\includegraphics[height=5.4cm,width=6.8cm]{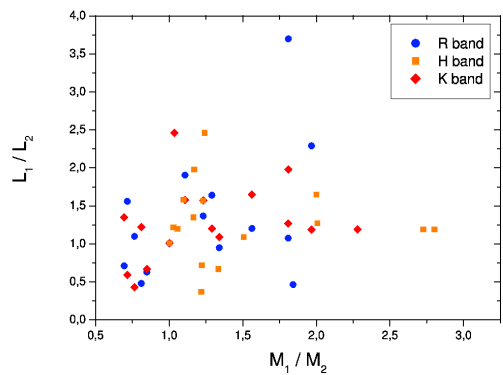}
\caption{Luminosity vs mass ratio of merging systems for R-band (circles), 
H-band (squares) and K-band (diamonds) data. The data are taken from 
\cite{kim02}, \cite{duc97}, and when not available, our H band images 
(boxes). This plot shows
the discrepancies between the luminosity estimates from different bands
and the fact that luminosity does not trace the mass in a robust way.  
\label{fig:lm}}
\end{figure*}

The R- and K-band luminosity ratios derived from the 
literature (Kim et al. 2002; Scoville et al. 2000; Duc et al 1997) 
are given in the same Table. The correlation inferred from 
Fig.~\ref{fig:lm} is rather weak, implying that tracing the mass content 
of each individual merger by its luminosity can be misleading, due to 
extinction and population effects. Further support for this argument comes
from the fact that, in several cases, the brightest nucleus in the NIR 
seems to be the faintest in the optical and vice-versa (see the luminosity 
ratios in Table~\ref{tab:ratio}). We conclude that the stellar kinematics 
are the most robust way to determine the mass ratios of merging 
galaxies. 

\begin{figure*}
\centering
\includegraphics[width=14.4cm]{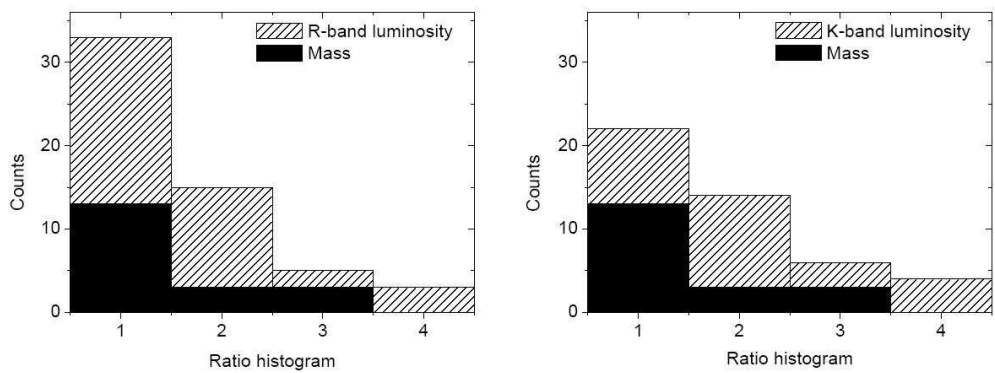}
\caption{Mass and luminosity ratio histogram. In filled bars we show the mass 
ratio of the ULIRGs in our sample, measured from the stellar kinematics. In 
shaded we show the R-band (left panel) and the K-band (right panel) luminosity 
ratio of the combined 1 Jy and Duc et al. (1997) samples. 
\label{fig:bins}}
\end{figure*}

In Fig.~\ref{fig:bins}, we place the binary systems (of luminosity ratio up 
to 4:1) of the combined samples of \cite{kim02} and \cite{duc97} in four 
luminosity ratio bins, for both the R-band (left panel) and the K-band (right
panel). The luminosity ratio distribution is different for the two bands due 
to extinction and population effects. We overplot our sample's mass ratio 
histogram in filled bars and we find that the distributions are
consistent, even though there are deviations in individual cases. This result 
implies that when the merging galaxies are nearly equal mass 
(i.e. the 1.5:1 ratio that we find for this
sample), individual deviations do not affect the statistical mean. 

\section{A model for the evolution of the mass ratio}
\label{sec:model}

We have run simulations of 1:1 and 3:1 mass ratio mergers of disk 
galaxies containing 10\% gas to test whether the mass ratio inferred from 
observations traces the intrinsic mass ratio of the galaxies and to quantify 
the influence of tidal effects and disk orientation. The disk galaxies and 
their orbits were set up in exactly the same way as in Naab \& Burkert (2003) 
(see their Section 2 and Table 1). To include the effects of a
dissipative component we replaced 10\% of the stellar mass in the initial
disks with isothermal gas at a temperature of approximately 10000 K.
The initial scale length $h$ of the stellar disk was equal to that of 
the gas disk. Each galaxy had a stellar bulge with 1/3 of the disk 
mass and was embedded in a pseudo-isothermal halo to guarantee a flat rotation 
curve at large radii. The gas disks were represented by 20000 SPH particles 
(6666 for the low mass disks) other particle numbers are as in Naab \& 
Burkert (2003). All galaxies approached each other on a nearly parabolic 
orbit with a pericenter distance of two disk scale lengths. The evolution of 
the stellar and the gas kinematics was computed with the N-body/SPH
code VINE using an isothermal equation of state for the gas. 

\begin{figure*}
\centering
\includegraphics[width=7.6cm]{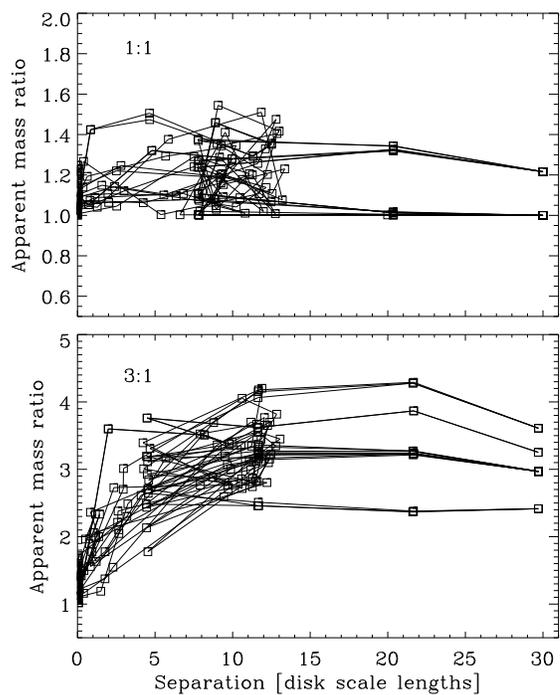}
\caption{ Apparent mass ratio measured as
$(\sigma_{1}^2 R_{\rm{eff},1})/(\sigma_{2}^2 R_{\rm{eff},2})$
versus distance of simulated merging disk galaxies with a true mass ratio
of 1:1 (upper panel) and 3:1 (lower panel). Every line represents one of 16
mergers with different initial disk orientations at a given mass ratio.
The squares indicate the measured mass ratios of the ongoing mergers
separated in time by a half-mass rotation period of the more
massive disk.\label{fig:model}}
\end{figure*}

In this paper
we analyzed mergers with 16 different initial disk orientations and mass ratios
1:1 and 3:1 (geometries 1-16 in Naab \& Burkert 2003, geometries 17-32 for the
3:1 mergers did not change the results presented here). We followed
every merger by analyzing snapshots in the orbital plane approximately every  
half-mass rotation period of the more massive disk. To avoid unrealistic 
values for $R_{\rm eff}$ when the galaxies overlap, we computed the 
effective radius of every galaxy as the projected spherical 
half-mass radius of the stellar particles within 5 scale lengths, taking into
account only particles of the galaxy itself. In addition, we computed the 
projected central stellar velocity dispersion for each galaxy within 
$0.5 R_{\rm eff}$ taking all stellar particles into account.
For each merger we have computed the time evolution of the apparent mass ratio
as $r_m=r_m{\rm (bulge)} = 
(\sigma_{1}^2 R_{\rm{eff,1}})/(\sigma_{2}^2 R_{\rm{eff,2}})$, where
the indices 1 and 2 declare the most and the least massive progenitor 
respectively. In Fig.~\ref{fig:model} we show the apparent mass ratios as a 
function of distance (in units of disc scale lenghts) for all 1:1 and 3:1 
merger remnants. 

Equal-mass mergers show apparent mass ratios in the range of $1 < r_m < 1.5$ 
which are very similar to the true mass ratio, independent of separation.
For 3:1 remnants, however, the scatter is larger and the apparent mass ratio is
in the range $1.5 < r_m < 4.3$ for distances greater than 10 scale lengths.
In particular, there is a trend for $r_m$ to decrease with decreasing distance 
which is mainly due to tidal heating of the low mass companion and not to a 
change in $R_{\rm eff}$. At distances below 5 scale lengths, a merger 
with an intrinsic mass ratio of 3:1 can easily be misclassified as 2:1. Given 
that the average half-light radius of this ULIRG sample is 2.2 kpc (and that
$R_{\rm eff} = 1.68 h$), 5 disk scale lengths equal 6.6 kpc. More than half of 
the mergers we observed  have a nuclear separation $<$ 6.6 kpc. Thus, the 
number of unequal-mass mergers that are able to lead to ultraluminous activity 
may be higher than what is measured. However, given that the 
majority ($\sim$ 60\%) of the sources are almost equal mass mergers, we do not 
expect the dynamical heating to drastically change our conclusions.


\section{Conclusions}
\label{sec:conc}
We have acquired spectroscopic H-band, long-slit data of 21 ULIRGs at a 
variety of prior to coalescence merger phases to study the mass ratios 
of the interacting objects that typically trigger ultraluminous activity.
Analysis of the kinematics indicates that the mean dispersion of
these ULIRGs is 142 \kms. The dynamical heating that occurs during
the merger leads to a low rotational component of the velocity compared to 
that of spirals, as the simulations of \cite{mihos00} predicted. The mean
inclination-corrected \vrot/$\sigma$ ratio of this sample is 0.77. The
mean mass ratio of the ULIRG progenitors is 1.5:1, which indicates that ULIRGs 
are mainly the products of almost equal mass mergers. Less frequently, 3:1
mergers appear in our sample. However, our simulations show that the 
unequal-mass merger categories may be undersampled due to dynamical 
heating and projection effects. We do not find significant evidence 
for minor mergers of progenitor mass ratio greater than 4:1; only one source, 
IRAS 10565+2448, appears as a minor merger in NIR images. However, the 
luminosity ratio of individual sources may significantly deviate from the 
actual mass ratio due to extinction and population effects. Using the stellar
dynamics is the most robust way to determine the mass content of a ULIRG. On 
a statistical basis,
the mass ratios implied by our kinematical analysis agree with the 1 Jy sample 
(R-band) luminosity ratios. The major mergers are typically those that are 
violent enough to drive an adequate amount of gas to the center of the 
system and trigger ultraluminous infrared bursts.

\acknowledgments

We are grateful to A. Verma for constructive comments. We thank 
A. Baker, M. Tecza, D. Rigopoulou, and C. Iserlohe for their input in the 
early phases of this study, and the ESO Paranal staff for 
their excellent support.


\appendix
\section{APPENDIX: Notes on individual sources}
\label{sec:individual}

IRAS 00199-0738: This object may be a multiple merger according to \cite{duc97}. Our spectroscopy shows that the nucleus to the north of the brightest nucleus 
is probably not at the same redshift, and does not belong to the same system. 
The radial distribution of the sources to the west and south-east also matches 
better that of a point-like (rather an extended) source which is broadened 
due to seeing. It is thus possible that none of these sources belongs 
to the particular merger.

IRAS 02364-4751: The spectroscopic results for this source indicate that
the difference in the dispersion of the two nuclei is 51 \kms. Given the
phase of the merger (nuclear separation of 1.5 kpc), one would expect 
smaller deviations in the progenitor dispersions which should
be closer to their common equilibrium value. However, the spectroscopic  
results show that the south nucleus has a large recession velocity with 
respect to its counterpart. This fact, combined with the proximity of the 
sources (that leads to a spatial overlap of the spectra of the two nuclei)
gives rise to an unrealistic increase of the southern nucleus dispersion. 

IRAS 10565+2448: A dwarf galaxy is seen at 6.5 kpc south-east of the bright 
nucleus, while the second nucleus of the merging system is at the north-east
of the bright nucleus (see Murphy et al. 1996).
The  $V_{\rm rot}$ value of this source is more 
likely between the inclination corrected and non-corrected one,
because it is in an early merger stage (nuclear separation of 21.7 
kpc), thus its $V_{\rm rot}$ and $V_{\rm rot} / \sigma$ ratio will 
probably be closer to those of a spiral.

IRAS 11095-0238: Recent H-band imaging obtained with the NICMOS camera at the
HST confirms that this is a close binary system (Veilleux et al.
2006, in preparation).

IRAS 12071-0444: This source was presented in \cite{tacconi02} as
a merged system. The data presented here (taken under better seeing
conditions) show the presence of two separate nuclei. HST NICMOS 
observations have also indicated the presence of two nuclei (Veilleux et al.
2006, in preparation).

IRAS 20046-0623: While two nuclei appear in the H-band images of this source,
IRAS 20046-0623 has often been treated as a single object in the literature 
due to the faintness of the eastern nucleus and the phase of the merger.
We have been able to extract the structural parameters and the spectroscopic 
results of the west, bright nucleus only.

IRAS 21329-2346: When deriving the H-band luminosity ratio of this system
from the (non-PSF subtracted) NICMOS images of Veilleux et al. (2006, 
in preparation), we find a luminosity ratio of 2.53, in good agreement with 
our results.


\clearpage

\begin{deluxetable}{ccccccc}
\tablecolumns{7}
\tabletypesize{\small}
\tablewidth{0pt}
\tablecaption{\label{tab:list} Binary ULIRGs source list}
\tablehead{
\colhead{Galaxy} & \colhead{RA}   & \colhead{Dec}    & \colhead{$z$} &
\colhead{log($L_{\rm IR}/$\hbox{L$_{\odot}$})} & \colhead{slit P.A.}   & 
\colhead{$t_{\rm integration}$}    \\

\colhead{(IRAS)} & \colhead{(2000)}   & \colhead{(2000)}    & \colhead{} &
\colhead{} & \colhead{(\degree)}   & \colhead{(mins)}    
}
\startdata
00199-7426 \tablenotemark{a}& 00:22:07.0 & -74:09:42 &  0.096 & 12.23 & -15,75,74 & 60,60,60 \\
01166-0844 & 01:19:07.6 & -08:29:10 &  0.118 & 12.03 & -60,29,29 & 60,60,60 \\
02364-4751 & 02:38:13.1 & -47:38:11 &  0.098 & 12.10 & 0,90 & 60,50 \\
06035-7102 & 06:02:54.0 & -71:03:10 &  0.0795 & 12.12 & 65,153,153 & 60,50,60\\
10190+1322 & 10:21:42   &  13:07:01 &  0.077 & 12.00 & 64,149,149 & 40,40,40 \\
10565+2448 & 10:59:18.1 &  24:32:34 &  0.0431 & 12.02 & -66,24 & 40,40 \\
11095-0238 & 11:12:03   & -02:54:18 &  0.106 & 12.20 & 39,129 & 120,120 \\
12071-0444 & 12:09:45.1 & -05:01:14 &  0.128  & 12.35 & -1,89 & 60,60 \\
12112+0305 & 12:13:47   &  02:48:34 &  0.073 & 12.28 & 37,99 & 60,60,40 \\
13335-2612 & 13:36:22   & -26:27:31 &  0.125 & 12.06 & -5 & 100 \\
13451+1232 & 13:47:33   &  12:17:23 &  0.122 & 12.28 & 104,13 & 80,120 \\ 
16156+0146 & 16:18:08   &  01:39:21 &  0.132 & 12.04 & -50,-51,40,40 & 60,60,60,60\\
16300+1558 & 16:32:20   &  15:51:49 &  0.242  & 12.63 & -1,89 & 150,90\\
19254-7245 & 19:31:21.4 & -72:39:18 &  0.0617 & 12.00 & -13,77 & 60,60\\
20046-0623 & 20:07:19.3 & -06:14:26 &  0.0844 & 11.97 & 69,159 & 60,60 \\ 
21130-4446 & 21:16:18.5 & -44:33:38 &  0.0926 & 12.02 & 33 & 40 \\
21208-0519 & 21:23:29 & -05:06:59 &  0.13   & 12.01 & -164,109,109 & 60,60,60\\
21329-2346 & 21:35:45   & -23:32:36 &  0.125  & 12.09 & 31 & 60 \\
22491-1808 & 22:51:49.2 & -17:52:23 &  0.0778 & 12.09 & -76,13,13 & 60,60,60 \\
23128-5919 & 23:15:46.8 & -59:03:15 &  0.045  & 11.96 & -5,84,84 & 40,40,40 \\
23234+0946 & 23:25:56.2 &  10:02:50 &  0.128  & 12.05 & -64,25 & 60,60 \\
\enddata

\tablenotetext{a}{This source may be a multiple merger (see the Appendix).}

\tablecomments{The coordinates, the redshift, the bolometric luminosity, as 
well as the slit positions and the respective integration time for our source
list are presented in this Table. }

\end{deluxetable}

\clearpage

\begin{deluxetable}{cccccc}
\tablecolumns{6}
\tabletypesize{\footnotesize}
\tablewidth{0pt}
\tablecaption{\label{tab:structure} ULIRG structural parameters}
\tablehead{
\colhead{Galaxy} & \colhead{$R_{\rm eff}$}   & \colhead{ellipticity}    
& \colhead{inclination} & \colhead{$\phi_{\alpha}$} & \colhead{nuclear 
separation} \\

\colhead{(IRAS)} & \colhead{(kpc)}   & \colhead{}    & \colhead{(\degree)} &
\colhead{(\degree)} & \colhead{(kpc)}  
}
\startdata
00199-7426    & 0.88 ($\pm0.04$) & 0.115 & 29 & 18 & \nodata \\
01166-0844(s) & 1.72 ($\pm1.23$) & 0.177 & 37 & -10 & 12.2 ($\pm$0.3) \\ 
01166-0844(n) & 1.55 ($\pm0.98$) & 0.178 & 37 & 61 & \nodata \\ 
02364-4751(s) & 1.45 ($\pm0.21$) & 0.250 & 44 & 52 & 1.6 ($\pm$0.3) \\ 
02364-4751(n) & 1.18 ($\pm0.14$) & 0.217 & 41 & -78 & \nodata \\
06035-7102(sw) & 1.79 ($\pm0.51$) & 0.331 & 51 & 39 & 10.4 ($\pm$0.2) \\
06035-7102(ne) & 1.41 ($\pm0.13$) & 0.398 & 57 & 34 & \nodata \\              
10190+1322(ne) & 1.43 ($\pm0.06$) & 0.298 & 48 & 6 & 6.5 ($\pm$0.2) \\ 
10190+1322(sw) & 2.40 ($\pm0.14$) & 0.223 & 41 & 37  & \nodata \\
10565+2448(s) & 0.79 ($\pm0.01$) & 0.042 & 17 & -84 & 23.3 ($\pm$0.1) \\ 
10565+2448(n) & 0.73 ($\pm0.10$) & 0.125 & 30 & 72 & \nodata \\
11095-0238(ne) & 2.07 ($\pm0.90$) & 0.151 & 34 & 27 & 3.8 ($\pm$0.3) \\ 
11095-0238(sw) & 3.04 ($\pm1.20$) & 0.398 & 57 & -22 & \nodata \\
12071-0444(s)  & 2.32 ($\pm1.05$) & 0.095 & 26 & -68 & 2.8  ($\pm$0.4) \\ 
12071-0444(n)  & 2.09 ($\pm0.70$) & 0.083 & 25 & 71  & \nodata  \\
12112+0305(sw) & 0.81 ($\pm0.01$) & 0.048 & 19 & 53 & 4.5 ($\pm$0.2) \\ 
12112+0305(ne) & 1.67 ($\pm0.29$) & 0.413 & 58 & 12 & \nodata \\
13335-2612(s) & 2.88 ($\pm0.17$) & 0.598 & 74 & -34 & 3.9 ($\pm$0.4) \\ 
13335-2612(n) & 2.25 ($\pm0.08$) & 0.098 & 27 & -55 & \nodata \\
13451+1232(w) & 2.59 ($\pm0.58$) & 0.094 & 26 & -1 & 5.3 ($\pm$0.3) \\
13451+1232(e) & 4.14 ($\pm2.16$) & 0.168 & 36 & -3 & \nodata       \\ 
16156+0146(n) & 0.90 ($\pm0.10$) & 0.128 & 31 & 89 & 8.8 ($\pm$1.2) \\ 
16156+0146(s) & 2.01 ($\pm0.12$) & 0.626 & 76 & 1 & \nodata \\
16300+1558(s) & 2.76 ($\pm1.37$) & 0.227 & 42 & -67 & 5.6 ($\pm$0.8) \\ 
16300+1558(n) & 4.83 ($\pm2.06$) & 0.351 & 53 & -14 & \nodata \\
19254-7245(s) & 0.97 ($\pm0.78$) & 0.288 & 47 & -21 & 10.2 ($\pm$0.2) \\ 
19254-7245(n) & 0.70 ($\pm0.35$) & 0.091 & 26 & 12 & \nodata \\
20046-0623(w)  & 2.67 ($\pm0.19$) & 0.673 & 82 & 6 & $<$4.4  \\  
20046-0623(e)  & \nodata & \nodata & \nodata & \nodata & \nodata \\
21130-4446(ne) & 1.71 ($\pm0.13$) & 0.398 & 57 & 8 & 5.4 $\pm$0.3)  \\
21130-4446(sw) & 2.69 ($\pm0.61$) & 0.584 & 72 & 32 & \nodata\\
21208-0519(s) & 3.66 ($\pm1.06$) & 0.139 & 32 & 21 & 17.9 ($\pm$0.4) \\
21208-0519(n) & 2.34 ($\pm0.67$) & 0.257 & 45 & -33 & \nodata\\ 
21329-2346(n) & 1.70 ($\pm0.12$) & 0.312 & 50 & -46 & 3.1 ($\pm$0.4) \\ 
21329-2346(s) & 1.42 ($\pm0.08$) & 0.128 & 31 & -88 & \nodata\\
22491-1808(e) & 1.99 ($\pm0.04$) & 0.370 & 54 & -62 & 3.3 ($\pm$0.2) \\ 
22491-1808(w) & 1.77 ($\pm0.10$) & 0.088 & 25 & 44 & \nodata\\
23128-5919(n) & 4.20 ($\pm0.08$) & 0.244 & 43 & -11 & 4.3  ($\pm$0.1) \\
23128-5919(s) & 4.16 ($\pm0.03$) & 0.296 & 48 & -7 & \nodata\\ 
23234+0946(n) & 2.12 ($\pm0.32$) & 0.154 & 34 & -32 & 9.4  ($\pm$0.4) \\  
23234+0946(s) & 3.28 ($\pm1.42$) & 0.116 & 29 & -58 & \nodata\\ 
\enddata
\tablecomments{ 
The ULIRG structural parameters are derived from the acquisition images.
For each system, the nuclear separation is given once and the nucleus 
with the most massive bulge appears first.}
\end{deluxetable}

\clearpage

\begin{deluxetable}{cccccccc}
\tablecolumns{8}
\tabletypesize{\footnotesize}
\tablewidth{0pt}
\tablecaption{\label{tab:velocities} Stellar velocities and resulting black 
hole masses}
\tablehead{
\colhead{Source} & \colhead{$\sigma$}   & 
\colhead{$V_{\rm rot}$(obs) \tablenotemark{a}}  &
\colhead{$V_{\rm rot}$ \tablenotemark{b}} &
\colhead{$V_{\rm rot}{(\rm obs)} / \sigma$} &
\colhead{$M_{\rm BH}$}   & 
\colhead{$M_{\rm BH}$(Edd.)}   & 
\colhead{$\eta_{\rm Edd}$}    \\

\colhead{(IRAS)} & \colhead{(km s$^{-1}$)}   & \colhead{(km s$^{-1}$)}    
& \colhead{(km s$^{-1}$)} & \colhead{} & \colhead{(\msun)} & \colhead{(\msun)}
& \colhead{} 
}
\startdata
00199-7426     & 137   ($\pm$ 55)  & 30 ($\pm$ 13)  & 76 & 0.22 
& $2.95 \times 10^7$ & $2.23 \times 10^7$ & 0.76 \\
01166-0844(s)  & 156   ($\pm$ 61)  &  \nodata  & \nodata  &  \nodata 
& $4.97 \times 10^7$ & $7.31 \times 10^6$ & 0.48 \\
01166-0844(n)  & 116   ($\pm$ 58)  &  \nodata  & \nodata  &  \nodata 
& $1.51 \times 10^7$ & $6.79 \times 10^6$ & 0.14 \\
02364-4751(s) & 151   ($\pm$ 32)  & \nodata  &  \nodata  & \nodata  
& $4.36 \times 10^7$ & $8.28 \times 10^6$ & 0.19 \\ 
02364-4751(n) & 100   ($\pm$ 32) & \nodata  &  \nodata  & \nodata  
& $8.32 \times 10^6$ & $8.28 \times 10^6$ & 0.96 \\ 
06035-7102(sw) & 136   ($\pm$ 24)  & 41 ($\pm$ 13)  & 52 & 0.30 
& $2.86 \times 10^7$ & $5.61 \times 10^6$ & 0.20 \\
06035-7102(ne) & 125   ($\pm$ 16)  & 14 ($\pm$ 15)  & 17 & 0.11 
& $2.04 \times 10^7$ & $1.17 \times 10^7$ & 0.57 \\
10190+1322(ne) & 169   ($\pm$ 35)  & 107 ($\pm$ 17) & 143 & 0.63 
& $6.86 \times 10^7$ & $7.69 \times 10^6$ & 0.11 \\
10190+1322(sw) & 127   ($\pm$ 12)  & 105 ($\pm$ 42) & 159 & 0.83 & 
$2.18 \times 10^7$ & $5.47 \times 10^6$ & 0.25 \\
10565+2448(s)  & 125   ($\pm$ 31)  & 134 ($\pm$ 23)  & 446 & 1.07 
& $2.04 \times 10^7$ & $1.38 \times 10^7$ & 0.68 \\
10565+2448(n)  &  \nodata & \nodata  & \nodata & \nodata  & \nodata & \nodata  
& \nodata  \\
11095-0238(ne) & 147   ($\pm$ 32)  & \nodata  &  \nodata  & \nodata  
& $3.92 \times 10^7$ & $1.04 \times 10^7$ & 0.27 \\ 
11095-0238(sw) & 137   ($\pm$ 38)  & \nodata  &  \nodata  & \nodata  
& $2.95 \times 10^7$ & $1.04 \times 10^7$ & 0.35 \\ 
12071-0444(s)  & 143   ($\pm$ 36)  & \nodata & \nodata & \nodata
& $3.50 \times 10^7$ & $1.32 \times 10^7$ & 0.38 \\
12071-0444(n)  & 130   ($\pm$ 29)  & \nodata & \nodata & \nodata
& $2.39 \times 10^7$ & $1.62 \times 10^7$ & 0.68 \\
12112+0305(sw) & 133   ($\pm$ 10)  & 34 ($\pm$ 19)  & 107 & 0.26 
& $2.62 \times 10^7$ & $1.53 \times 10^7$ & 0.58 \\
12112+0305(ne) & 124   ($\pm$ 23) &  5 ($\pm$ 18)  & 6 & 0.04  
& $1.98 \times 10^7$ & $9.81 \times 10^6$ & 0.50 \\
13335-2612(s)  & 175   ($\pm$ 43)  &  \nodata  & \nodata  & \nodata   
& $7.89 \times 10^7$ & $8.25 \times 10^6$ & 0.10 \\
13335-2612(n)  & 140   ($\pm$ 27)  &  \nodata  & \nodata &  \nodata  
& $3.22 \times 10^7$ & $6.86 \times 10^6$ & 0.21 \\
13451+1232(w)  & 167   ($\pm$ 48)  &  \nodata & \nodata &  \nodata   
& $6.54 \times 10^7$ & $1.75 \times 10^7$ & 0.27 \\
13451+1232(e)  & 146   ($\pm$ 28)  &  \nodata  & \nodata  & \nodata   
& $3.81 \times 10^7$ & $7.57 \times 10^6$ & 0.20 \\
16156+0146(n)  & 189   ($\pm$ 27)  &  \nodata  & \nodata &  \nodata   
& $1.08 \times 10^8$ & $9.22 \times 10^6$ & 0.09 \\
16156+0146(s)  & \nodata & \nodata & \nodata & \nodata & \nodata 
& $5.21 \times 10^6$ & \nodata \\
16300+1558(s)  & 141   ($\pm$ 47)  & \nodata &  \nodata & \nodata 
& $3.31 \times 10^7$ & $2.81 \times 10^7$ &  0.85\\ 
16300+1558(n)  &  \nodata  & \nodata  & \nodata & \nodata & \nodata & \nodata
 & \nodata \\ 
19254-7245(s)  & 175   ($\pm$ 24)  & 99 ($\pm$ 22)  & 135 & 0.57 
& $7.89 \times 10^7$ & $9.16 \times 10^6$ & 0.12 \\
19254-7245(n)  & 120   ($\pm$ 19)  & 47 ($\pm$ 34)  & 113  & 0.39 
& $1.73 \times 10^7$ & $4.00 \times 10^6$ & 0.23 \\
20046-0623(w)  & 145   ($\pm$ 14)  & 103 ($\pm$ 13) & 104 & 0.71 
& $3.71 \times 10^7$ & $1.23 \times 10^7$ & 0.33 \\
20046-0623(e)  &  \nodata & \nodata  & \nodata & \nodata  & \nodata & \nodata  
& \nodata  \\
21130-4446(ne) & 165   ($\pm$ 37)  &  \nodata &  \nodata & \nodata   
& $6.23 \times 10^7$ & $8.48 \times 10^6$ & 0.14 \\
21130-4446(sw) & 152   ($\pm$ 28)  &  \nodata & \nodata &  \nodata   
& $4.48 \times 10^7$ & $5.33 \times 10^6$ & 0.12 \\
21208-0519(s)  & 171   ($\pm$ 22)  & \nodata  & \nodata & \nodata   
& $7.19 \times 10^7$ & $4.28 \times 10^6$ & 0.06 \\
21208-0519(n)  & 126   ($\pm$ 21)  & \nodata & \nodata & \nodata   
& $2.12 \times 10^7$ & $9.19 \times 10^6$ & 0.44 \\  
21329-2346(n)  & 115   ($\pm$ 21)  & \nodata  & \nodata & \nodata   
& $1.46  \times 10^7$ & $ 1.13 \times 10^7$ & 0.78 \\
21329-2346(s)  & 113   ($\pm$ 22)  & \nodata  & \nodata & \nodata   
& $1.36  \times 10^7$ & $ 4.86 \times 10^6$  & 0.36 \\
22491-1808(e)  & 146   ($\pm$ 20)  & 16 ($\pm$ 31)  & 20 & 0.11 
& $3.81 \times 10^7$ & $7.02 \times 10^6$ & 0.18 \\
22491-1808(w)  & 121   ($\pm$ 34)  & 27 ($\pm$ 50)  & 64 & 0.23 
& $1.79 \times 10^7$ & $9.17 \times 10^6$ & 0.51 \\
23128-5919(n)  & 151   ($\pm$ 21)  & 29 ($\pm$ 16)  & 43 & 0.20 
& $4.36 \times 10^7$ & $4.13 \times 10^6$ & 0.09 \\
23128-5919(s)  & 148   ($\pm$ 18)  & 82 ($\pm$ 12)  & 110 & 0.56 
& $4.02 \times 10^7$ & $7.87 \times 10^6$ & 0.20 \\
23234+0946(n)  & 152   ($\pm$ 23)  & \nodata  & \nodata & \nodata   
& $4.48 \times 10^7$ & $1.16 \times 10^7$ & 0.26 \\
23234+0946(s)  & 113   ($\pm$ 40)  & \nodata  & \nodata &  \nodata   
& $1.36 \times 10^7$ & $3.14 \times 10^6$ & 0.23 \\
\enddata

\tablenotetext{a}
{Velocity corrected for deviations from the major axis of rotation.}
\tablenotetext{b}
{Observed velocity corrected for inclination effects.}

\tablecomments{ 
The stellar dispersion and rotational velocities, and the $V_{\rm rot}/\sigma$ 
ratio are derived from the spectra of Fig.~\ref{fig:spectra} with the aid of 
the parameters of Table ~\ref{tab:structure}. The dynamical and Eddington 
black hole mass of each nucleus and the ratio of the two are also presented 
here.}

\end{deluxetable}

\clearpage

\begin{deluxetable}{ccccccc}
\tablecolumns{7}
\tabletypesize{\small}
\tablewidth{0pt}
\tablecaption{\label{tab:ratio} Progenitor mass ratios}
\tablehead{
\colhead{Galaxy} & \colhead{$r_m$(bulge) \tablenotemark{a}}   & 
\colhead{$r_m$ \tablenotemark{b}} & 
\colhead{$r_m$(aperture) \tablenotemark{c}} &
\colhead{apert. ('')} & \colhead{$r_L$(R band)} & \colhead{$r_L$(K band)} 
}
\startdata
IRAS 01166-0844  & 2.01 & 2.01 & 1.81 & 1.18 & 0.70 & 1.08 \\
IRAS 02364-4751  & 2.80 & 2.80 & 2.28 & 1.18 & \nodata & \nodata \\
IRAS 06035-7102  & 1.07 & 1.03 & 0.81 & 1.47 & 1.21 & 0.48 \\
IRAS 10190+1322  & 0.95 & 1.16 & 0.69 & 1.47 & 1.74 & 0.71 \\
IRAS 11095-0238  & 1.28 & 1.28 & 0.87 & 0.88 & \nodata & \nodata \\
IRAS 12071-0444  & 1.34 & 1.34 & 1.21 & 1.03 & \nodata & \nodata \\
IRAS 12112+0305  & 1.48 & 1.22 & 0.72 & 1.47 & 1.56 & 0.59 \\
IRAS 13335-2612  & 2.00 & 2.00 & 1.56 & 1.47 & 1.03 & 1.20 \\
IRAS 13451+1232  & 1.22 & 1.22 & 0.76 & 1.76 & 1.10 & 0.43 \\
IRAS 14348-1447  & 1.05 & 1.06 & 1.29 & \nodata & 1.26 & 1.64 \\
IRAS 19254-7245  & 2.94 & 2.73 & 1.97 & 1.18 & 1.45 & 2.29 \\
IRAS 21130-4446  & 1.33 & 1.33 & 0.84 & 1.32 & 0.76 & 0.63 \\
IRAS 21208-0519  & 2.88 & 2.88 & 1.84 & 1.47 & 1.32 & 0.47 \\
IRAS 21329-2346\ & 1.24 & 1.24 & 1.04 & 0.88 & \nodata  & \nodata \\
IRAS 22491-1808  & 1.64 & 1.51 & 1.34 & 1.04 & 1.10 & 0.95 \\
IRAS 23128-5919  & 0.95 & 1.10 & 1.11 & 2.35 & 1.20 & 1.91 \\ 
IRAS 23234+0946  & 1.17 & 1.17 & 1.81 & 1.76 & 2.38 & 3.70 \\ 
arp 220          & \nodata & \nodata & 1.23 & \nodata &  \nodata & 1.37 \\
NGC 6240         & \nodata & \nodata & 1.00 & \nodata &  \nodata & 1.01 \\
\enddata

\tablenotetext{a}{Bulge mass ratio calculated using the stellar dispersion 
and the effective radius of each progenitor.}
\tablenotetext{b}{Total mass ratio calculated using the stellar dispersion, 
rotational velocity (whenever possible), and effective radius of each 
progenitor.}
\tablenotetext{c}{Total mass ratio calculated within a constant aperture, 
common for both progenitors.}

\tablecomments{ 
The total baryonic mass ratio at the half-light-radius, and at a specific 
aperture (given in the fourth column), the bulge mass ratio and the R and K 
luminosity ratios of the progenitors can be found in this Table.
}
\end{deluxetable}

\clearpage


\begin{figure*}
\includegraphics[height=20cm,width=16cm]{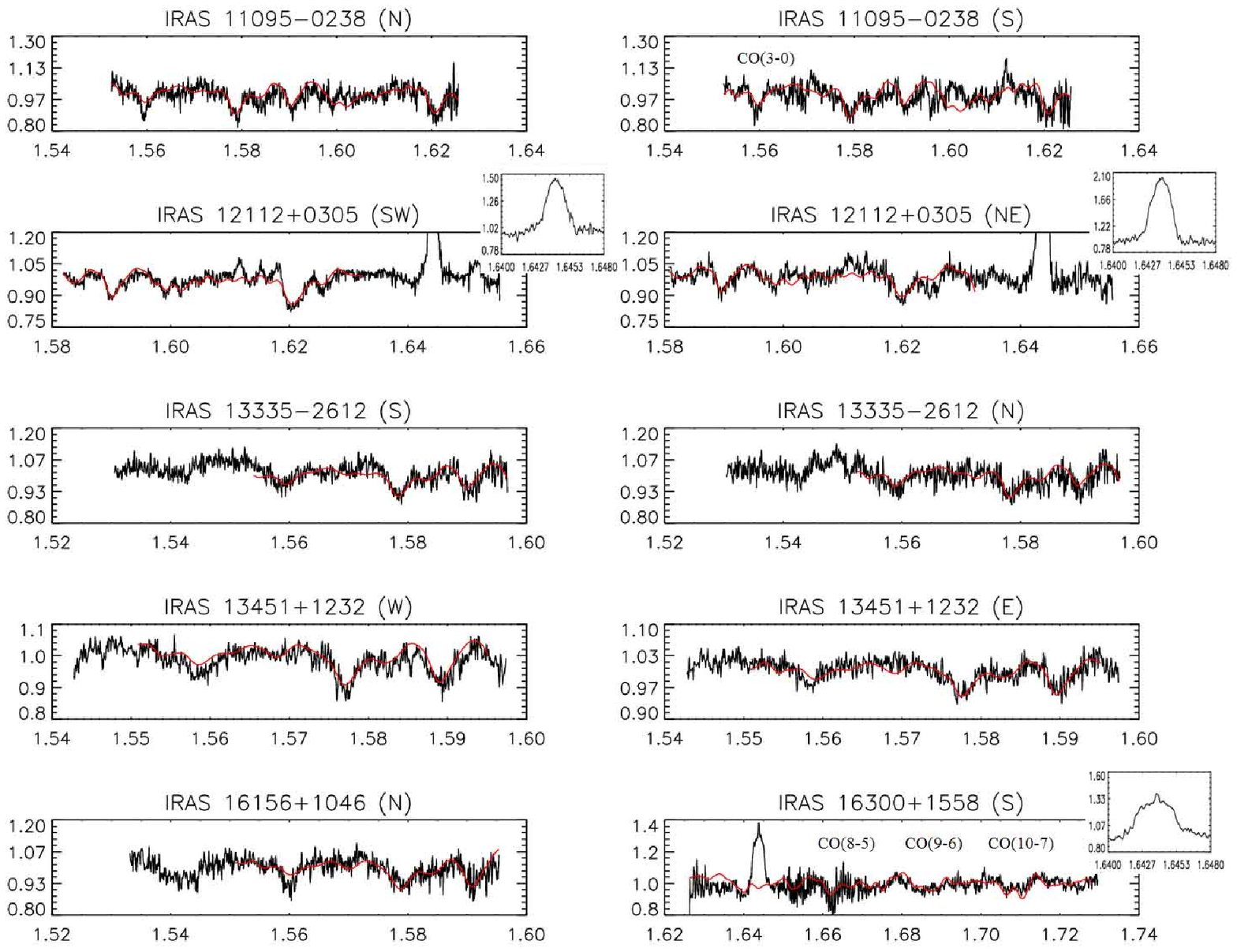}
\begin{center}
Fig.~\ref{fig:spectra} continued.
\end{center}
\end{figure*}


\begin{figure*}
\includegraphics[height=20cm,width=16cm]{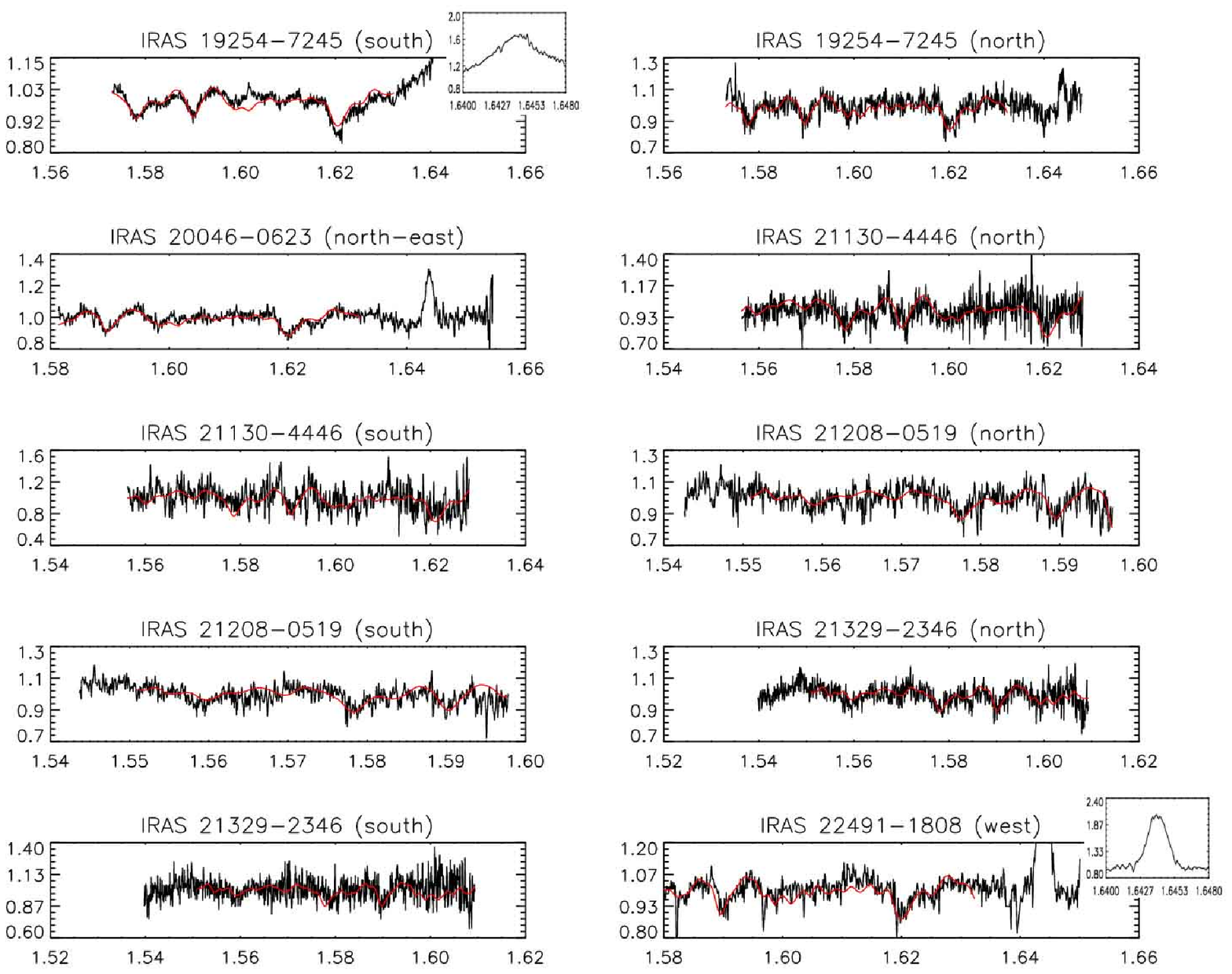}
\begin{center}
Fig.~\ref{fig:spectra} continued.
\end{center}
\end{figure*}


\begin{figure*}
\includegraphics[height=20cm,width=16cm]{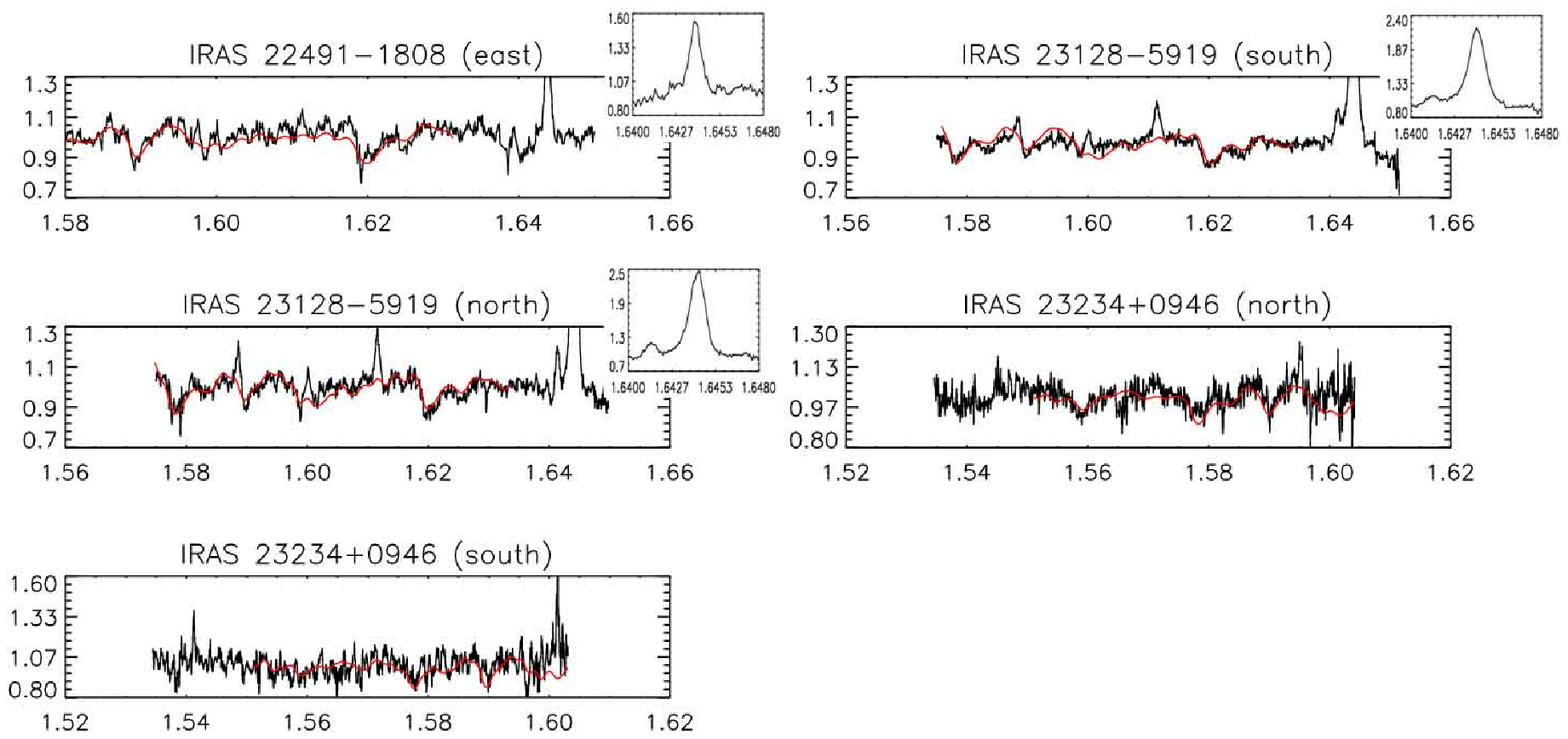}
\begin{center}
Fig.~\ref{fig:spectra} continued.
\end{center}
\end{figure*}

\clearpage

\begin{figure*}
\includegraphics[width=16cm]{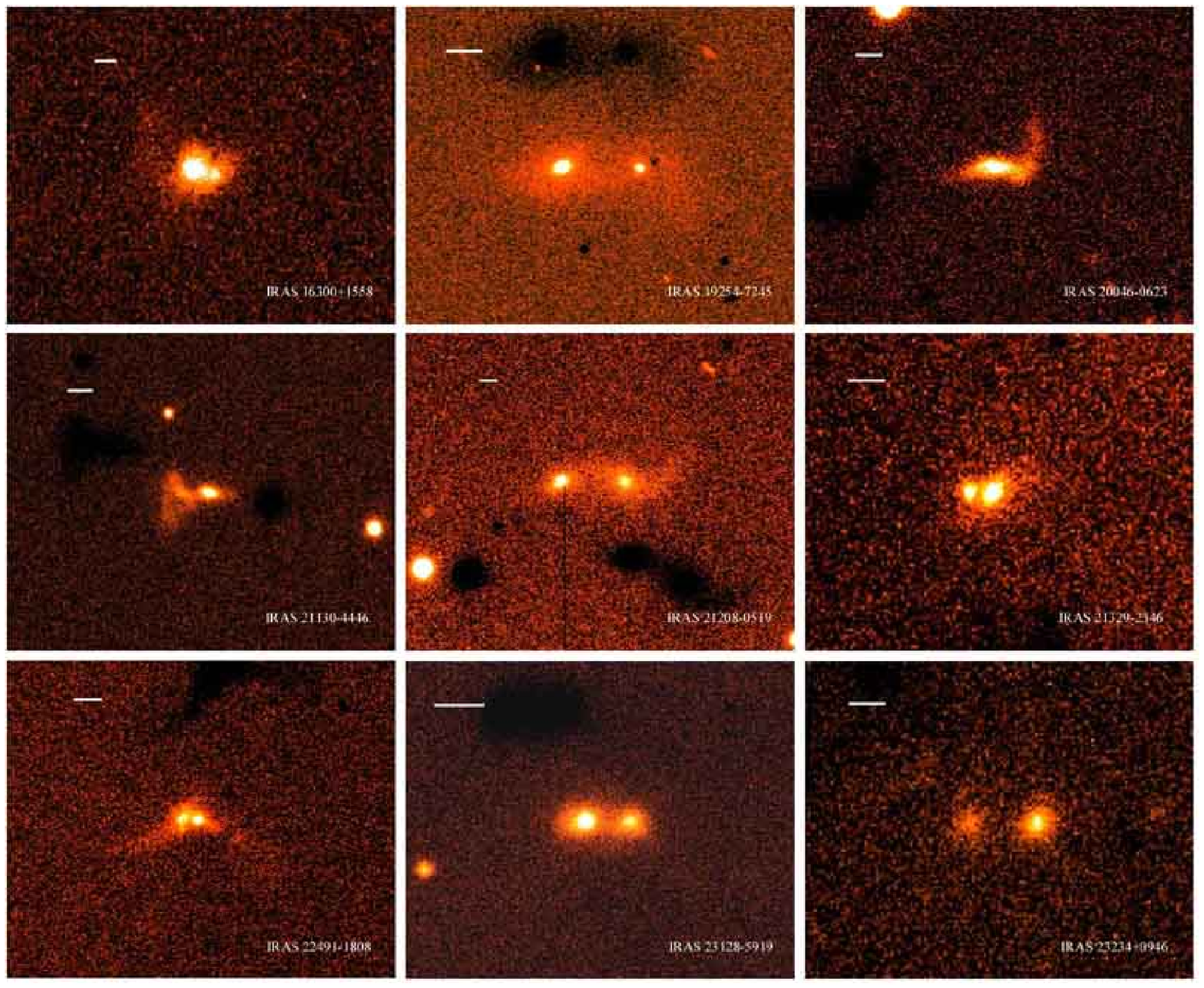}
\begin{center}
Fig.~\ref{fig:acquisition} continued.
\end{center}
\end{figure*}

\clearpage

\end{document}